\documentclass[pra,twocolumn,showpacs,amssymb,amsmath]{revtex4}
\def\>{\rangle}
\def\<{\langle}
\def\Tr{\hbox{Tr}}
\usepackage{graphicx}
\begin{document}

\title{Phase-covariant cloning of coherent states}
\author{Massimiliano F. Sacchi} \homepage{http://www.qubit.it}  
\affiliation{QUIT Quantum Information Theory Group, 
CNISM and CNR - Istituto Nazionale per la Fisica della Materia,
  Dipartimento di Fisica ``A. Volta'',  University of Pavia, 
via A. Bassi 6, I-27100 Pavia, Italy,\\
Istituto Universitario di Studi Superiori (IUSS), via Luino 4, I-27100
Pavia, Italy.} 
\date{\today}

\begin{abstract} We consider the problem of phase-covariant cloning
  for coherent states. We show that an experimental scheme based on
  ideal phase measurement and feedforward outperforms the
  semiclassical procedure of ideal phase measurement and preparation
  in terms of fidelity. A realistic scheme where the ideal phase
  measurement is replaced with double-homodyne detection is shown to
  be unable to overcome the semiclassical cloning strategy. On the
  other hand, such a realistic scheme is better than semiclassical
  cloning based on double-homodyne phase measurement and preparation.
\end{abstract}


\maketitle 

\section{Introduction} 

The impossibility of exact quantum cloning---i.e. copying the state of
a quantum system to a larger number of uses \cite{noclon}---has
stimulated the search for quantum devices that can emulate cloning
with the highest possible fidelity. Many optimal cloners have been
found, since the simplest case of qubits \cite{buzhill,gismass,bruss},
including general finite-dimensional systems \cite{werner}, restricted
sets of input states \cite{darlop,darmacc}, and infinite-dimensional
systems such as harmonic oscillators---the so called continuous
variable cloners \cite{cerf,cerfbrauns}.  More recently, cloning for
mixed states has been considered, for both qubits \cite{prl} and
Gaussian continuous variable systems \cite{cv1}.

In the present paper, we analyze the cloning of coherent states by
phase-preserving maps. More precisely, this means that we will
consider a set of coherent states with a fixed known absolute value of
the amplitude and unknown phase, thus studying cloning machines that
are covariant with respect to the phase Abelian group $U(1)$, instead
of the Weyl-Heisenberg group of complex displacements, as done for
arbitrary coherent states. Clearly, one expects to improve the
fidelity of the $N$ to $M$ cloning of coherent states
$F_{NM}=\frac{MN}{MN+M-N}$ \cite{cerfbrauns}, since a restricted set
of states is considered.

The issue of quantum cloning is studied for both its fundamental
interest and its relevance in practice, in the realm of quantum key
distribution and quantum cryptography \cite{key,key2}. In particular,
the study of phase-covariant cloning may be of interest in analyzing
the security of quantum cryptographic protocols based on phase coding.
While extensive work has been devoted to phase-covariant cloning for
qubits and qudits \cite{pc1,pc2,pc3,pc4,pc5,pc6,pc7,pc8}, the case of
coherent states is still unexplored.

The general theory for the problem of cloning that is covariant with
respect to a given group of transformations has been given in
\cite{darlop}. However, the optimal phase-covariant cloning for
coherent states has never been derived. The main reason is that the
phase-covariance property entails infinite equivalent representations,
corresponding to Hilbert subspaces with fixed photon-number
difference, which makes the problem highly untractable.  The
derivation of the optimal phase-covariant cloning would be interesting
to give a benchmark for realizable cloning schemes, but presumably the
result would be almost academic, beyond the experimental capability of
the best quantum optics laboratories.

Here, we will study the problem of phase-covariant cloning of coherent
states by considering experimental schemes based on splitting the
input coherent states, phase measurement, and feedforward.
Measurement-and-feedforward schemes have been considered recently for
cloning of coherent states and Gaussian mixed states \cite{ula,cv1},
where optimality has also been proved.  We will compare such
experimental schemes with the semiclassical cloning procedure based on
phase estimation and preparation of coherent states.

The paper is organized as follows.  In Sec. II we sketch the general
problem for phase-covariant cloning.  In Sec. III we evaluate the
average fidelity of semiclassical cloning based on measurement and
preparation, considering both ideal phase measurement and
double-homodyne detection.  In Sec. IV we consider two cloning schemes
based on splitting, phase measurement and feedforward, when the ideal
phase measurement and the phase measurement achieved by marginal of
double-homodyne detection are adopted. Clearly, for a realistic
implementation, the second scheme is more interesting, and in fact the
effect of nonunit quantum efficiency of detectors is also taken into
account. In these schemes two parameters can be set free: the ratio of
the splitting and the feedforward gain, and one can suitably optimize
over them to maximize the fidelity. In Sec. V, we compare the
semiclassical strategy with the feedforward schemes. The following
result is obtained: when an ideal phase measurement is performed, the
feedforward scheme outperforms the semiclassical method, whereas when
double-homodyne detection is adopted---a much simpler and viable
measurement---the feedforward scheme is in practice unable to overcome
the semiclassical cloning, even for very high quantum efficiency.
However, the realistic scheme is better then the corresponding
semiclassical cloning based on double-homodyne detection and
preparation.  In Sec. VI we draw  conclusions with closing remarks.

\section{Optimal phase-covariant cloning} 

The problem of optimal phase-covariant cloning for continuous variable
systems can be posed as follows. Let us denote by $a$ ($a^\dag $) the
annihilation (creation) operator of a single-mode radiation field,
with $[a,a^\dag]=1$. We are given a set of $N$ input copies $(e^{i
  a^\dag a \phi }\rho _0 e^{-i a^\dag a\phi })^{\otimes N}$, where
$\rho _0$ is known, and the phase $\phi \in [0,2\pi ]$ is unknown, and
we are interested in looking for a physical transformation (i.e. a
completely positive map $\cal E$) which produces $M$ output copies
with maximum fidelity with respect to the input, independently of the
value of the unknown phase $\phi $.  The fidelity of the output clones
will then be the same for any input phase.

The input-output description in terms of completely positive maps is
equivalent \cite{bds} to a scheme where an ancilla is added, a unitary
transformation is performed, and the ancilla is discarded (i.e. traced
out). The latter scheme is the traditional approach to the problem of
optimal quantum cloning \cite{buzhill}.

The optimal map can be searched among the covariant ones that satisfy
\begin{eqnarray} 
&&{\cal E}[(e^{i a^\dag a \phi }\rho _0 e^{-i a^\dag
a \phi })^{\otimes N}] \nonumber \\ 
= &&  (e^{i a^\dag a \phi })^{\otimes M} {\cal E}(\rho _0
 ^{\otimes N}) (e^{-i a^\dag a \phi })^{\otimes M} \;.
\label{covp}
\end{eqnarray} 
The covariance property
leads to a striking simplification of the problem \cite{darlop}, since one can
exploit the theory of irreducible representation of groups. However,
in the phase-covariant problem for continuous variables, the optimal
solution has never been reported, since infinite equivalent
representations appear when considering the phase Abelian group $U(1)$. 
In fact, it is useful to consider
  the Choi-Jamio\l kowski bijective correspondence of {\em completely
  positive} (CP) maps $\cal E$ from the input Hilbert space ${\cal H}_\mathrm{in}$ 
to the output Hilbert space 
  ${\cal H}_\mathrm{out}$ and {\em positive} operators $ R_{\cal E}$ acting on
  ${\cal H}_\mathrm{in}\otimes {\cal H}_\mathrm{out}$, which is given by the
  following expressions
\begin{equation} \begin{split}
  &R_{\cal E}=I \otimes {\cal E} (|\Omega \>\< \Omega |)\;,\\
  &{\cal E}(\rho)=\Tr_\mathrm{in}[(\rho ^\tau \otimes I_\mathrm{out})
  R_{\cal E}]\;, \end{split} 
\end{equation} 
where
  $|\Omega\>=\sum_{n=0}^{\infty}|\psi_n\>|\psi_n\>$ is a maximally
  entangled vector of ${\cal H}_\mathrm{in} ^{\otimes 2}$, and $\tau $
  denotes the transposition on the fixed basis $\{|\psi_n\>\}$. In terms of
  the operator $R_{\cal E}$ the covariance property (\ref{covp}) can be
  written as 
\begin{eqnarray} [R_{\cal E},
(e^{-i a^\dag a \phi}) ^{\otimes
  N} \otimes (e^{i a^\dag a \phi
  })^{\otimes M}
]=0\;,\qquad \forall \phi \;.\label{com} \end{eqnarray} 
The trace-preserving condition of the map $\cal E$ is given in terms of
  the positive operator $R_{\cal E}$ by the constraint 
\begin{eqnarray}
\Tr
  _\mathrm{out} [R_{\cal E}]=I_\mathrm{in}\;,
\label{tpc}\end{eqnarray}
 and the 
global fidelity of the cloning machine is given by 
\begin{equation}
F=\Tr [(\rho ^{\tau \otimes N}\otimes \rho ^{\otimes M})\, R_{\cal E}
]\;.\label{fopt}
\end{equation}
The optimal cloning machine can be obtained by maximizing $F$ in Eq. 
(\ref{fopt}) under the constraints (\ref{com}) and (\ref{tpc}). 
The solution of such a problem is presently unknown. 

\section{Semiclassical phase-covariant cloning} 
The semiclassical strategy for cloning quantum states consists in
performing a quantum measurement on the given input copies, and
creating suitable output copies according to the measurement outcome.
Then, for phase-covariant cloning, one can perform the ideal phase
measurement to infer the phase of the unknown input with best
accuracy.

The quantum estimation of the phase 
is the essential problem of high sensitive
interferometry, and has received much attention in quantum optics
\cite{rev1}. For a single-mode electromagnetic field, the measurement
cannot be achieved exactly, even in principle, due to the lack of a
unique self-adjoint operator \cite{london}. 

The most general and concrete approach to the problem of the phase
measurement is quantum estimation theory \cite{helstrom}. 
Quantum estimation theory provides a more
general description of quantum statistics in terms of POVM's (positive
operator-valued measures) and gives the theoretical definition of an
optimized phase measurement. The most powerful method for deriving the
optimal phase measurement was given by Holevo \cite{Holevo} in the
covariant case.  In this way the optimal POVM for phase estimation has
been derived for a single-mode field.  More generally, the problem of
estimating the phase shift has been addressed in Ref. \cite{chiara}
for any degenerate shift operator with arbitrary discrete spectrum. 

\par The optimal
POVM for the phase measurement can be written 
in terms of projectors on
Susskind-Glogower states \cite{ssg} as follows 
\begin{eqnarray}
d\mu (\phi )=\frac {d\phi}{2\pi }|e^{i\phi } \rangle \langle
e^{i\phi } |\;,\label{sg}
\end{eqnarray}
where $|e^{i\phi }\rangle =
\sum_{n=0}^{\infty} e^{i\phi n}|n \rangle $, and $\{ | n\rangle \}$ denotes the Fock basis.  
Notice that the states $|e^{i\phi }\rangle $ are not normalizable,
neither are they orthogonal; however, they provide a resolution of the identity, and
thus guarantee the completeness of the POVM, namely,
\begin{eqnarray}
\int _{0}^{2\pi }d\mu (\phi )=I\;.
\end{eqnarray}
For a system in state $\rho$, the POVM in Eq. (\ref{sg}) gives the
ideal phase distribution $p(\phi )$ according to Born's rule 
\begin{eqnarray}
p(\phi )=\hbox{Tr}[d\mu (\phi )\,\rho]= 
\frac {d\phi}{2\pi }\langle e^{i\phi } |\rho |
e^{i\phi } \rangle \;.\label{pfi}
\end{eqnarray}

For a set of $N$ input coherent states $|\alpha \rangle ^{\otimes N}$, 
the semiclassical strategy consists in concentrating by a  multisplitter 
the $N$ coherent states into a single coherent state with amplitude $\sqrt N \alpha $, 
performing the ideal phase measurement, and creating $M$ coherent states 
$|e^{i \phi } |\alpha |\rangle $, where $\phi $ is the phase-measurement outcome. 
Hence, on average one obtains the $M$ output copies 
\begin{eqnarray}
&& {\cal E}(|\alpha \rangle \langle \alpha |^{\otimes N}) 
\nonumber \\  = && 
  \int _{0}^{2\pi }\frac{d\phi }{2\pi }\, |e^{i \phi }|\alpha |\rangle \langle 
e^{i\phi} |\alpha || ^{\otimes M} \,|\langle \sqrt 
N \alpha | e^{i\phi }\rangle |^2 \;.
\end{eqnarray}
Using the relations
\begin{eqnarray}
|\alpha \rangle =e^{-\frac{|\alpha |^2}{2}}\sum _{n=0}^\infty 
\frac{\alpha ^n}{\sqrt {n !}}|n \rangle \,,\qquad  
|\langle \alpha |\beta \rangle |^2=e^{-|\beta -\alpha|^2} 
\;,
\end{eqnarray}
we can evaluate the fidelity of each of the $M$ copies with the initial 
input as follows
\begin{eqnarray}
F_{cl}^{SG} &=&
  \int _{0 }^{2\pi }\frac{d\phi }{2\pi }\, 
|\langle \alpha |e^{i \phi } |\alpha |\rangle |^2
\,|\langle \sqrt {N} \alpha | e^{i\phi }\rangle |^2   \\
&=&  \int _{0}^{2\pi }\frac{d\phi }{2\pi }\, e^{-|\alpha |^2[2(1-\cos \phi)+N]}
 \nonumber \\ & \times & 
\sum_{n,m=0}^\infty \frac{(\sqrt{N}|\alpha |)^{n+m}}{\sqrt{n! m!}}
\,e^{i\phi (n-m)}
\nonumber \\ &
= &e^{-|\alpha |^2 (2+N)}
\sum_{n,m=0}^\infty \frac{(\sqrt{N}|\alpha |)^{n+m}}{\sqrt{n! m!}}\,I_{n-m}(2|\alpha |^2), \nonumber
\label{fcl}
\end{eqnarray}
where we used the identity
\begin{equation}
\int _{0} ^{2\pi } \frac{d\phi }{2\pi }\,e^{z \cos \phi +i m\phi }=I_m (z)\;, 
\label{bes} 
\end{equation}
$I_m (z)$ denoting the $m$-th order modified Bessel function.

In a realistic semiclassical cloning strategy, the ideal phase
measurement is beyond the available technology \cite{idphase}.  Thus,
we consider the case where double-homodyne detection is used instead,
in order to estimate the phase.  The probability distribution for the
phase measurement of coherent states by means of double-homodyne
detection with quantum efficiency $\eta $ is given in the Appendix.
Using this result, the fidelity is then given by
\begin{eqnarray}
F_{cl}^{DH} &=& e^{-|\alpha |^2 (2+N)}
\sum_{n,m=0}^\infty \frac {\Gamma \left (\frac {n+m}{2}+1 \right )}
{n! m!} \nonumber \\&\times  & \left (
\frac{\sqrt{N}|\alpha | }
{\sqrt{\Delta ^2 _\eta +1}}\right )^{n+m}\,
I_{n-m}(2|\alpha |^2)\;.
\label{fcl2}
\end{eqnarray}
where $\Delta ^2 _\eta =\frac{1-\eta }{\eta }$. 

As for any measure-and-prepare scheme, notice that the fidelities
(\ref{fcl}) and (\ref{fcl2}) do not depend on the number of output
copies $M$. The schemes are clearly covariant, i.e. the fidelity
depends just on $|\alpha |$, and not on the phase $\arg \alpha$.

\section{Phase-covariant cloning via partial measurement and
feedforward} 
In this Section we consider the following covariant
cloning strategy based on phase measurement and feedforward. First,
the $N$ coherent states are concentrated by a multisplitter into a
single coherent state with amplitude $\sqrt N \alpha $. Second, this 
state is split into two coherent states by a beam splitter with tunable
transmissivity, thus giving two coherent states $|\sqrt N \alpha \cos
\theta \rangle $ and $|\sqrt N \alpha \sin \theta \rangle $, where
$\theta \in [0,\pi ]$. Third, a phase measurement described by a POVM
$d\mu (\phi)$ is performed on the second coherent state $|\sqrt N
\alpha \sin \theta \rangle $, and the displacement $D(k |\alpha |
e^{i\phi })$ is the feedforward input to the first coherent state $|\sqrt N
\alpha \cos \theta \rangle $ , where $\phi $ is the measurement
outcome and $k$ a free parameter. Finally, this  coherent state is
split into $M$ coherent states whose amplitude is then rescaled by
$\sqrt M $.  By averaging over the phase-measurement outcomes, such a
cloning strategy produces the following $M$ copies: 
\begin{eqnarray}
&& {\cal E}(|\alpha \rangle \langle \alpha |^{\otimes N})  
=    \int _{0}^{2\pi }  \frac{d\phi }{2\pi } 
\nonumber
\\  &\times  &
\left | \frac{\sqrt N \alpha \cos \theta + k 
|\alpha |e^{i \phi }}{ \sqrt M } \right \rangle \left \langle 
\frac{\sqrt N \alpha \cos \theta + k 
|\alpha |e^{i \phi }}{ \sqrt M } \right 
|^{\otimes M}
\nonumber \\
& \times  & 
\langle \sqrt N \alpha \sin \theta  |d\mu (\phi )
|\sqrt N \alpha \sin \theta \rangle \;.
\end{eqnarray}
The fidelity of each of the $M$ copies with the initial input depends on the parameters $\theta $ and $k$, along with the POVM $d\mu (\phi) $, and can be written as follows
\begin{eqnarray}
F_{ff} &= &
  \int _{0}^{2\pi } \frac{d\phi }{2\pi } \,
| \langle \alpha |(\sqrt N \alpha \cos \theta + k 
|\alpha |e^{i \phi })/ \sqrt M \rangle |^2 
\nonumber \\& \times  & 
\langle \sqrt N \alpha \sin \theta  |d\mu (\phi )
|\sqrt N \alpha \sin \theta \rangle \;.\label{fff}
\end{eqnarray}

\subsection{Using ideal phase measurement}
With a similar derivation of Eq. (\ref{fcl}), 
when the ideal phase measurement (\ref{sg}) is adopted the 
fidelity in Eq. (\ref{fff}) is given by 
\begin{eqnarray}
F_{ff}^{SG}&=&e^{-\frac{|\alpha |^2 }{M}
[(\sqrt M -\sqrt N \cos\theta )^2+NM \sin^2 
\theta +k^2]}
\nonumber \\
& \times & 
\sum_{n,m=0}^\infty \frac{(\sqrt{N}|\alpha |\sin\theta
  )^{n+m}}{\sqrt{n! m!}}
\nonumber \\
& \times & 
I_{n-m}\left[\frac{2|\alpha |^2}{M}(\sqrt M -\sqrt N \cos \theta )k\right ].  
\label{fffsg}
\end{eqnarray}

\subsection{Using double-homodyne detection}
Using Eq. (\ref{app}) of the Appendix for the 
phase measurement of coherent states
 by means of double-homodyne detection, 
one obtains the following expression for the 
fidelity of the cloning based on double-homodyne detection and feedforward:
\begin{eqnarray}
F_{ff}^{DH} &= &e^{-\frac{|\alpha |^2 }{M}[(\sqrt M -\sqrt N \cos\theta
)^2+ \frac{NM \sin^2 \theta }{\Delta _\eta ^2 +1}+k^2]} \nonumber \\ &
\times & \sum_{n,m=0}^\infty \frac{(\sqrt{N}|\alpha |\sin\theta
)^{n+m}} {\sqrt{\Delta ^2 _\eta +1}}\,\frac {\Gamma \left (\frac
{n+m}{2}+1 \right )} {n! m!}
\nonumber \\
& \times & 
 I_{n-m}\left[\frac{2|\alpha
|^2}{M}(\sqrt M -\sqrt N \cos \theta )k\right ].  \label{fffdh}
\end{eqnarray}

\section{Numerical results} 
The expression of the fidelities for the
feedforward-based schemes can be optimized versus the parameters
$\theta $ and $k$, namely the beam splitter transmissivity and the gain
factor of the displacement. In this Section we present some numerical results of 
the optimized fidelities, and compare the quality of the cloning with 
the semiclassical one.  

\par In Fig. 1 we report the fidelity for 1-to-2 phase-covariant
cloning of coherent states versus the absolute amplitude $|\alpha |$:
the 
feedforward scheme with ideal phase measurement (solid line), and
double-homodyne phase measurement with quantum efficiency $\eta =1$
(dashed line) and $\eta =0.8$ (dotted line); semiclassical cloning via ideal
phase measurement and preparation (dot-dashed line).

\begin{figure}[htb]
\begin{center}
\includegraphics[scale=1]{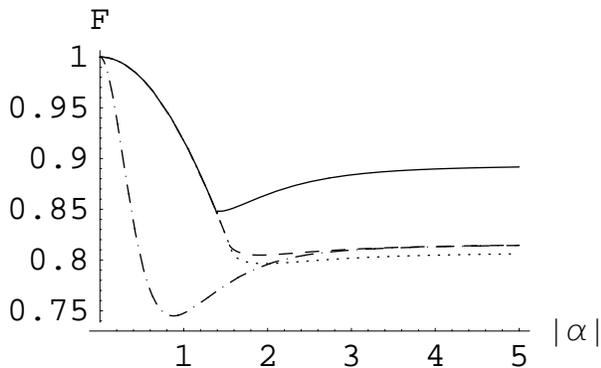}
\caption{Fidelity for 1-to-2 phase-covariant cloning of coherent
states versus the absolute amplitude $|\alpha |$: the feedforward scheme
with ideal phase measurement (solid line), and double-homodyne phase
measurement with quantum efficiency $\eta =1$ (dashed line) and $\eta =0.8$
(dotted line); semiclassical cloning via ideal phase measurement and
preparation (dot-dashed line).}  \label{f:fig1} \end{center} \end{figure}

It turns out that 
 when the ideal phase measurement is performed, the feedforward 
  scheme outperforms the semiclassical method, whereas when
  double-homodyne detection is adopted 
the feedforward scheme is in practice  unable to do better than the 
semiclassical cloning, even for very high quantum
  efficiency.  We notice that for low values of $|\alpha |$,
  i.e. $|\alpha | \lesssim 1.6$, the optimization of the feedforward
  schemes gives $\theta = k = 0$, which means that cloning is achieved
  just by splitting the coherent states, where the phase is
  preserved. This can be easily understood since a coherent state with
  such a low intensity has a very poor phase resolution due to the
  high fidelity with the vacuum state, and hence a cloning strategy
  based on phase-measurement has a poor efficiency.  On the other
  hand, for both the semiclassical and the feedforward cloning
  schemes, the fidelity rapidly saturates to a constant for $|\alpha |
  \geq 4$, independently of the amplitude. In all cases, as expected,
  the fidelity overcomes the value $2/3$ of the cloning for arbitrary
  coherent states.  For $N$ to $M$ cloning, the above features are
  essentially unchanged.

\par In Fig. 2 we show the values of the fidelity for 1-to-M
phase-covariant cloning with $|\alpha |=5$ versus the output number of
copies $M$, for the feedforward-scheme with both ideal phase
measurement (gray) and double-homodyne phase measurement with quantum
efficiency $\eta =1$ (black). For increasing number of output copies,
even the value of the fidelity of the feedforward scheme and ideal
phase measurement tends to the classical value $F \simeq 0.8157 $,  
confirming that quantum information becomes 
classical when distributed to many users \cite{chir}.

\begin{figure}[htb]
\begin{center}
\includegraphics[scale=1]{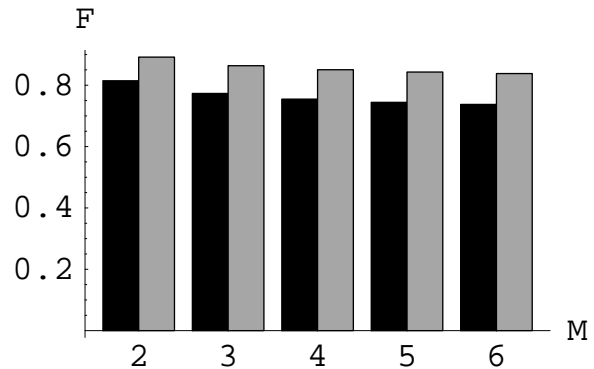}
\caption{Fidelity for 1-to-M phase-covariant cloning of coherent states with 
absolute amplitude $|\alpha |=5$ versus the output number of copies $M$: 
feedforward-scheme with ideal phase measurement 
(gray), and double-homodyne phase measurement with quantum efficiency $\eta =1$ 
(black).}
\label{f:fig2}
\end{center}
\end{figure}

In Tab. 1, we report the explicit values for the fidelity of different
$1$-to-$2$ cloning schemes and for different values of $|\alpha |$, along with
the optimal splitting parameter $\theta $ and gain parameter $k$ for
the feedforward schemes. Although a realistic scheme based on 
double-homodyne detection and feedforward does not 
  outperform the semiclassical cloning (i.e. $F_{ff}^{DH} < F_{cl}^{SG}$), 
however it does outperform semiclassical cloning based on double-homodyne
  detection and preparation (i.e. $F_{ff}^{DH} > F_{cl}^{DH}$).

\section{Conclusions}
We have studied the problem of phase-covariant cloning of coherent
states. After showing the difficulty of the problem of deriving the
optimal phase-covariant cloning, we have compared cloning schemes
based on phase measurement and feedforward with semiclassical schemes
based on phase measurement and preparation. In both cases, we
considered an ideal phase measurement and double-homodyne phase
measurement.  In the feedforward schemes we have shown that the
fidelity can be optimized over the splitting parameter of the input
copies and the feedforward gain.  When the ideal phase measurement is
performed, the feedforward scheme outperforms the semiclassical
method, whereas when double-homodyne detection is adopted the
feedforward scheme is unable to overcome the semiclassical cloning,
even for very high quantum efficiency. In any case, the realistic
scheme is better then the corresponding semiclassical cloning based on
double-homodyne detection and preparation.

\begin{table}[hbt]
\begin{center}
\begin{tabular}{|c|ccc|ccc|c|c|}
\hline
 $|\alpha |$ & $F_{ff}^{SG}$ & $ k $ & 
 $\theta $ & $F_{ff}^{DH\ (\eta =0.8)} $ 
& $k$ & $\theta $ & 
$F_{cl}^{SG}$ & 
$F_{cl}^{DH}$  \\ 
\hline
3 & 0.884 & 0.746 & 0.861 & 0.802 & 0.697 & 0.809 & 0.810 &  
 0.703 \\ 
\hline
4 & 0.890 & 0.722 & 0.818 & 0.805 & 0.701 & 0.861 & 0.813 &  
 0.705 \\
\hline
5 & 0.892 & 0.714 & 0.802 & 0.806 & 0.703 & 0.793 & 0.814 &  
 0.706 \\
\hline
6 & 0.893 & 0.711 & 0.796 & 0.807 & 0.704 & 0.791 & 0.815 &  
 0.706 \\ \hline 
\end{tabular} \end{center} 
\caption[fake]{Numerical
 values for the fidelity of $1$-to-$2$ phase-covariant cloning of
 coherent states versus the absolute value of the amplitude $|\alpha
 |$ for different schemes: feedforward and ideal phase measurement
 ($F_{ff}^{SG}$); feedforward and double-homodyne phase measurement with quantum efficiency $\eta =0.8$ 
 ($F_{ff}^{DH}$); ideal phase measurement and preparation
 ($F_{cl}^{SG}$); double-homodyne phase measurement and preparation
 ($F_{cl}^{DH}$).  For the feedforward schemes the optimized values of
 the gain parameter $k$ and the splitting parameter $\theta $ are also
 reported.}  
\label{tab}\end{table}

\par Our results show that an experimental verification of the quantum
features of a phase-covariant cloning of coherent states is very
demanding, since a feedforward scheme would require an ideal phase
measurement, which {\em per se} is a hard task. Unfortunately, a phase
measurement based on the customary double-homodyne detection is too 
noisy to produce good clones of coherent states.

\section*{Appendix}

The POVM describing double-homodyne detection (or, equivalently,
heterodyne detection) with quantum efficiency $\eta $ is given by
\cite{ex}
\begin{equation} 
d\mu (\alpha )=\frac {d^2 \alpha }{\pi }\int _{\mathbb
C}\frac{d^2\beta }{\pi \Delta ^2 _\eta }\, e^{-\frac {|\beta -\alpha
|^2}{\Delta ^2 _\eta }} |\beta \rangle \langle \beta |\;,
\end{equation} 
where $\Delta ^2_\eta =\frac{1-\eta }{\eta }$, and $d^2
\alpha = d \hbox{Re} \alpha \,d \hbox{Im} \alpha =|\alpha | \,d|\alpha
|\,d \phi $ with $\alpha =|\alpha | e^{i \phi }$.  The probability
distribution $p(\phi |\gamma)$ for the phase measurement on a coherent
state $|\gamma \rangle $ is then obtained as a marginal integration
over the modulus $|\alpha |$ of the measurement outcome $\alpha $, and
one has 
\begin{eqnarray} p(\phi |\gamma ) &=& \frac{d\phi }{\pi }
\int _{0}^\infty
d|\alpha |\, |\alpha | \int _{\mathbb C}\frac{d^2\beta }{\pi \Delta ^2
_\eta }\, e^{-\frac {|\beta -\alpha |^2}{\Delta ^2 _\eta }} |\langle
\gamma |\beta \rangle |^2 
\nonumber \\ & = & 
\frac{d\phi }{\pi }\int _{0}^\infty d|\alpha
|\, |\alpha | \int _{\mathbb C}\frac{d^2\beta }{\pi \Delta ^2 _\eta
}\, e^{-\frac {|\beta -\gamma |^2}{\Delta ^2 _\eta }} e^{-|\beta
-\alpha |^2} 
\nonumber \\ & = &
\frac{d\phi }{\pi (\Delta ^2 _\eta +1)}\int
_{0}^\infty d|\alpha |\, |\alpha | e^{-\frac {|\alpha -\gamma
|^2}{\Delta ^2 _\eta +1 }} 
\nonumber \\ &= & 
\frac{d\phi }{2 \pi }
e^{-\frac {|\gamma
|^2}{\Delta ^2 _\eta +1 }} \int _{0}^\infty dt
\,e^{-t}\,e^{\sqrt {\frac{t}{\Delta ^2 _\eta +1}} (e^{i\phi }\gamma ^*
+e^{-i\phi }\gamma )} \nonumber \\ & =&\frac {1}{2\pi }\,e^{-\frac
{|\gamma |^2}{\Delta ^2 _\eta +1 }} \,\sum_{n,m=0}^\infty \frac
{\Gamma \left( \frac {n+m}{2}+1 \right )}{n!m!} 
\nonumber \\& \times & 
\frac{\gamma
^{*n}{\gamma ^m}}{(\Delta ^2 _\eta +1)^{\frac{n+m}{2}}}\, e^{i\phi
(n-m)}
\;,\label{app} \end{eqnarray} 
where $\Gamma (x) =
\int_{0}^\infty dt\,e^{-t}\,t^{x-1}$ denotes the usual gamma function.

\section*{Acknowledgments} 
This work has been sponsored by Ministero Italiano dell'Universit\`a e
della Ricerca (MIUR) through FIRB (2001) and PRIN 2005.


\begin{thebibliography}{99}
\bibitem{noclon} W. K. Wootters and W. H.  Zurek, {\em Nature} {\bf 299}, 802 (1982); D.~Dieks,
  Phys. Lett. A, {\bf 92}, 271 (1982); H. P. Yuen, Phys. Lett. A {\bf 113}, 405 (1986); G. C.
  Ghirardi and T. Weber, Nuovo Cimento B {\bf 78}, 9 (1983). 
\bibitem{buzhill} V. Bu\v zek and M. Hillery, Phys. Rev. A {\bf 54},
  1844 (1996).
\bibitem{gismass} N. Gisin and S. Massar, Phys. Rev. Lett. {\bf 79},
  2153 (1997).
\bibitem{bruss} D. Bruss, D. P. DiVincenzo, A. Ekert, C. A. Fuchs, C.
  Macchiavello, and J. A. Smolin, Phys.  Rev. A {\bf 57}, 2368 (1998).
\bibitem{werner} R. F. Werner, Phys. Rev. A {\bf 58}, 1827 (1998); R.
  F. Werner and M. Keyl, J. Math. Phys. {\bf 40}, 3283 (1999).
\bibitem{darlop} G. M. D'Ariano and P. Lo Presti, Phys. Rev. A {\bf
    64}, 042308 (2001).
\bibitem{darmacc} G. M. D'Ariano and C. Macchiavello, Phys. Rev. A
  {\bf 67}, 042306 (2003).
\bibitem{cerf} N. J. Cerf, A. Ipe, and X. Rottenberg, Phys. Rev. Lett.
  {\bf 85}, 1754 (2000); N. J. Cerf and S. Iblisdir, Phys. Rev. Lett.
  {\bf 87}, 247903 (2001).
\bibitem{cerfbrauns} S. L. Braunstein, N. J. Cerf, S. Iblisdir, P. van
  Loock, and S. Massar, Phys. Rev. Lett. {\bf 86}, 4938 (2001).
\bibitem{prl} G. M. D'Ariano, C. Macchiavello, and P. Perinotti, Phys.
  Rev. Lett. {\bf 95}, 060503 (2005).
\bibitem{cv1}G. M. D'Ariano, P. Perinotti, and M. F. Sacchi, New J. Phys. 
{\bf 8}, 99 (2006); Europhys. Lett. {\bf 75}, 195 (2006); quant-ph/0609046. 
\bibitem{key}C. H. Bennett and G. Brassard, in {\em Proceedings of the
    IEEE International Conference on Computers, Systems, and Signal
    Processing, Bangalore, India} (IEEE, New York, 1984), p.  175; C.
  H. Bennett, Phys. Rev. Lett. {\bf 68}, 3121 (1992); N. Gisin, G.
  Ribordy, W. Tittel, and H. Zbinden, Rev. Mod. Phys. {\bf 74}, 145
  (2002).
\bibitem{key2}A. K. Ekert, Phys. Rev. Lett. {\bf 67}, 661 (1991);
  Nature {\bf 358}, 14 (1992).  
\bibitem{pc1}D. Bruss, M. Cinchetti, G. M. D'Ariano, and C. Macchiavello, 
Phys. Rev. A {\bf 62}, 012302 (2000). 
\bibitem{pc2}V. Karimipour and A. T. Rezakhani, Phys. Rev. A {\bf 66}, 
052111 (2002). 
\bibitem{pc3} H. Fan, K. Matsumoto, X. B. Wang, and M. Wadati, 
Phys. Rev. A {\bf 65}, 012304 (2001). 
\bibitem{pc4} H. Fan, H. Imai, K. Matsumoto, and X. B. Wang, 
Phys. Rev. A {\bf 67}, 022317 (2003). 
\bibitem{pc5}J. Fiur\'a\v sek, Phys. Rev. A {\bf 67}, 052314 (2003). 
\bibitem{pc6} F. Buscemi, G. M. D'Ariano, and C. Macchiavello, Phys. Rev. 
A {\bf 71}, 042327 (2005). 
\bibitem{pc7}F. Sciarrino and F. De Martini, Phys. Rev. A {\bf 72}, 062313 
(2005). 
\bibitem{pc8} F. Buscemi, G. M. D'Ariano, C. Macchiavello, and P. Perinotti, 
Phys. Rev. A {\bf 74}, 042309 (2006).
\bibitem{ula}U. L. Andersen,  V. Josse, and 
  and G. Leuchs, Phys. Rev. Lett. {\bf 94}, 240503 (2005).
\bibitem{bds} F. Buscemi, G. M. D'Ariano, and M. F. Sacchi, 
Phys. Rev. A {\bf 68}, 042113 (2003).
\bibitem{rev1} Physica Scripta T {\bf 48} (1993) (special issue on {\em
Quantum Phase and Phase Dependent measurements}).
\bibitem{london} F. London, Z. Phys. {\bf 37}, 915 (1926); {\bf
  40}, 193 (1927).
\bibitem{helstrom} C. W. Helstrom, {\em Quantum Detection 
and Estimation Theory} (Academic, New York, 1976). 
\bibitem{Holevo} A. S. Holevo. {\em Probabilistic and statistical
aspects of quantum theory} (North-Holland, Amsterdam, 1982).
\bibitem{chiara} G. M. D'Ariano, C. Macchiavello, and M. F. Sacchi,
  Phys. Lett. A {\bf 248}, 103 (1998).  
\bibitem{ssg} L. Susskind and
  J. Glogower, Physics {\bf 1}, 49 (1964).  
\bibitem{idphase}A large
  class of theoretical schemes, however, have been proposed in
  F. Buscemi, G. M. D'Ariano, and M. F. Sacchi, Phys. Lett. A {\bf
  312}, 315 (2003).  
\bibitem{chir}G. Chiribella and G. M. D'Ariano, 
Phys. Rev. Lett. {\bf 97}, 250503 (2006).
\bibitem{ex}See, for example, G. M. D'Ariano,
  M. G. A. Paris, and M. F. Sacchi, {\em Quantum Tomography}, in
  Advances in Imaging and Electron Physics {\bf 128}, 205-308 (2003),
  quant-ph/0302028.  
\end{thebibliography}
 \end{document}